\documentclass[12pt]{article}
\usepackage{authblk}
\usepackage[bookmarksnumbered, colorlinks, plainpages]{hyperref}
\usepackage{amsmath, amsthm, amscd, amsfonts, amssymb, graphicx, color, booktabs}
\usepackage{siunitx}
\usepackage{braket}

\usepackage{graphicx}
\usepackage{subcaption}
\usepackage{placeins}

\usepackage[bottom]{footmisc}

\numberwithin{equation}{section}
\definecolor{email}{rgb}{0.00,0.00,0.84}
\begin{document}
\setcounter{page}{1}

\title{\large \bf 12th Workshop on the CKM Unitarity Triangle\\ Santiago de Compostela, 18-22 September 2023 \\ \vspace{0.3cm}
\LARGE Updates on inclusive charmed and bottomed meson decays from the lattice}

\author{Ryan Kellermann\textsuperscript{1}, Alessandro Barone\textsuperscript{2,3,4}, Shoji Hashimoto\textsuperscript{1}, Andreas J\"uttner\textsuperscript{3,4,5}, Takashi Kaneko\textsuperscript{1,6} \\
        {\small \textsuperscript{1}Theory Center, Institute of Particle and Nuclear Studies, High Energy Accelerator Research Organization (KEK), Tsukuba 305-0801, Japan and School of High Energy Accelerator Science, The Graduate University for Advanced Studies (SOKENDAI), Tsukuba 305-0801, Japan} \\ 
        {\small \textsuperscript{2}PRISMA+ Cluster of Excellence \& Institut f\"ur Kernphysik, Johannes-Gutenberg-Universit\"at Mainz, D-55099 Mainz, Germany} \\ 
        {\small \textsuperscript{3}School of Physics and Astronomy, University of Southampton, Southampton SO17 1BJ, United Kingdom} \\
        {\small \textsuperscript{4}STAG Research Center, University of Southampton, Southampton SO17 1BJ, UK} \\
        {\small \textsuperscript{5}CERN, Theoretical Physics Department, Geneva, Switzerland} \\
        {\small \textsuperscript{6}Kobayashi-Maskawa Institute for the Origin of Particles and the Universe, Nagoya University, Aichi 464–8602, Japan} \\ }

\maketitle

\begin{abstract}

With the development of lattice QCD in recent years, a determination of decay rates for inclusive semileptonic decays from lattice correlators now seems viable. We report on the calculation of the inclusive semileptonic decay of the $D_s$ meson on the lattice. We simulate the $D_s \rightarrow X_s\ell\nu_\ell$ process with M\"obius domain-wall charm and strange quarks, whose masses were approximately tuned to the physical values. We cover the whole kinematical region. We present a general overview on how the inclusive decays are treated on the lattice and discuss ongoing challenges, such as the estimation of systematic errors and future prospects of the project.

\end{abstract} \maketitle

\section{Introduction}

In recent years, experiments have revealed a puzzling tension in B-decays, namely, in the determination of the CKM parameters $|V_{ub}|$ and $|V_{cb}|$ from exclusive and inclusive methods~\cite{Workman:2022ynf}. This discrepancy provides an opportunity for theorists and experimentalists to improve their understanding of these decays. Furthermore, the search for new physics requires precise theoretical predictions from the Standard Model. In view of these points, recently, ideas to extend the application of lattice QCD towards the description of inclusive decays have been proposed~\cite{Gambino:2020crt, Gambino:2022dvu, Hansen:2017mnd, Hansen:2019idp, Bulava:2021fre}. These approaches utilize either the Chebyshev approximation or the Backus-Gilbert approach to obtain the energy integral of the hadronic tensor, which defines the inclusive decay rates. In this paper, we do not address the analysis strategy used to analyze inclusive decays from lattice correlators but instead only present preliminary results obtained from the analysis as well as strategies towards estimating systematical errors. We refer to \cite{Gambino:2020crt, Gambino:2022dvu, Kellermann:2022mms} for an overview and \cite{Barone:2023tbl} for a comparison between the Chebyshev and Backus-Gilbert approaches. While a direct application of this method towards the determination of the the aforementioned CKM matrix elements $|V_{ub}|$ and $|V_{cb}|$ would be desirable, the treatment of the bottom sector in lattice QCD comes with its own set of challenges. To be precise, these concern the challenge of simulating the bottom mass at its physical value due to the large discretization errors expected for $m_b$. These errors are under better control for the charm sector, leading us to first apply the method towards the decay of charmed mesons.

The rest of this paper is structured as follows. We briefly review the inclusive semileptonic decay on the lattice in Sec. \ref{sec:InclusiveLattice}. In Sec. \ref{sec:SystematicErrors} we discuss the systematic errors associated with the inclusive analysis. Finally, in Sec. \ref{sec:Future}, we provide a short summary and discuss the future direction of this project as well as possibe applications.

\section{Inclusive semileptonic decays on the lattice}
\label{sec:InclusiveLattice}

The total decay rate of the inclusive semileptonic decay is written as
\begin{align}
    \Gamma \sim \int_0^{\pmb{q}^2_{\text{max}}} d\pmb{q}^2 \sqrt{\pmb{q}^2} \sum_{l=0}^{2} \bar{X}^{(l)}(\pmb{q}^2) \, ,
    \label{equ:TotalDecayRate}
\end{align}
where $\bar{X}^{(l)}(\pmb{q}^2)$ contains the integral over the hadronic final state energy $\omega$
\begin{align}
    \begin{split}
    \bar{X}_\sigma^{(l)}(\pmb{q}^2) &= \int_{\omega_0}^{\infty} d\omega \, W^{\mu\nu}(\pmb{q},\omega) e^{-2\omega t_0} K^{(l)}_{\mu\nu, \sigma} (\pmb{q},\omega) \\
    &= \braket{\psi^\mu(\pmb{q})|K^{(l)}_{\mu\nu, \sigma}(\pmb{q},\hat{H})|\psi^\nu(\pmb{q})} \, ,
    \end{split}
\end{align}
with the hadronic tensor $W^{\mu\nu}(\pmb{q},\omega)$, $\ket{\psi^{\nu}(\pmb{q})} = e^{-\hat{H} t_0} \tilde{J}^{\nu}(\pmb{q}, 0) \ket{D_s} / \sqrt{2M_{D_s}}$ and $\tilde{J}^{\nu}(\pmb{q}, 0)$ being the Fourier transformed currents. The lower limit  $0 \leq \omega_0 \leq \omega_{\text{min}}$ can be chosen freely as there are no states below the lowest lying energy state $\omega_{\text{min}}$. The parameter $t_0$ is introduced to avoid the contact term which receives contributions from the opposite time ordering corresponding to unphysical states. In the definition of $\bar{X}_\sigma^{(l)}(\pmb{q}^2)$ above 
\begin{align}
   K^{(l)}_{\mu\nu, \sigma}(\pmb{q},\omega) = e^{2\omega t_0} \sqrt{\pmb{q}^2}^{2-l} (m_{D_s} - \omega)^l \theta_{\sigma}(m_{D_s} - \sqrt{\pmb{q}^2} - \omega) \, ,
   \label{equ:KernelFunction}
\end{align}
defines the \textit{kernel function} and $\theta_\sigma(x)$ is a sigmoid function with smearing width $\sigma$.

On the lattice we compute
\begin{align}
    C_{\mu\nu}(t) = \frac{1}{2M_{D_s}} \braket{D_s|\tilde{J}^{\mu\dagger}(\pmb{q},0) e^{-\hat{H}t} \tilde{J}^{\nu}(\pmb{q},0)|D_s} \, ,
    \label{equ:FourPointCorrelator}
\end{align}
and the calculation of the inclusive decay rate is reduced to the one of finding an appropriate polynomial approximation of the kernel function $K^{(l)}_{\mu\nu, \sigma}(\pmb{q},\hat{H})$.

We employ the shifted Chebyshev polynomials $\tilde{T}_j(x)$, with $x = e^{-\omega}$ and define the approximation as
\begin{align}
  \braket{K_{\mu\nu, \sigma}^{(l)}} \simeq \frac{1}{2} \tilde{c}_{0}^{(l)} \braket{\tilde{T}_0}_{\mu\nu} + \sum_{k=1}^{N} \tilde{c}_{k}^{(l)} \braket{\tilde{T}_k}_{\mu\nu} \, .
  \label{equ:ChebApprox}
\end{align}
Here, $\tilde{c}_k^{(l)}$ are analytically known coefficients and $\braket{\tilde{T}_k}$ are referred to as \textit{Chebyshev matrix elements}. We use the notation $\braket{\cdot} \equiv \braket{\psi^\mu|\cdot|\psi^\nu}/\braket{\psi^\mu|\psi^\nu}$. For simplicity, we skip the indices $\mu, \nu$ going forward.

The matrix elements are extracted from a fit to the correlator data following
\begin{align}
    \bar{C}(t) = \sum_{j=0}^{t} \tilde{a}_j^{(t)} \braket{\tilde{T}_j} \, ,
    \label{equ:FitCorrelator}
\end{align}
where $\tilde{a}_j^{(t)}$ are obtained from the power representation of the Chebyshev polynomials, see (A.24) and (A.25) of \cite{Barone:2023tbl} for the definition of $\tilde{a}_j^{(t)}$, and $\bar{C}(t)$ is constructed from \eqref{equ:FourPointCorrelator} as $\bar{C}(t) = C(t+2t_0)/C(2t_0)$. To maximize the available data we choose $t_0 = 1/2$.
We use priors to ensure that the fitted Chebyshev matrix elements satisfy the condition that the Chebyshev polynomials are bounded, i.e. $\left|\braket{\tilde{T}_j}\right| \leq 1$. We refer to \cite{Barone:2023tbl} for more details on the Chebyshev approximation and the practical application.

\section{First numerical results}
\label{sec:Numerics}

\begin{figure}[bt!]
    \centering
    \includegraphics[width=0.7\textwidth]{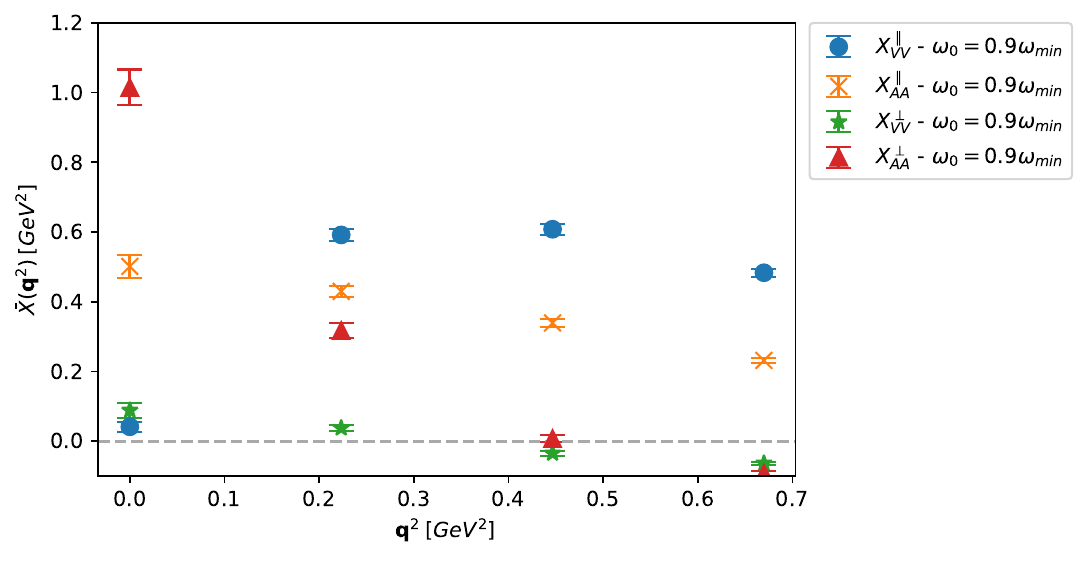}
    \caption{Contributions to the total $\bar{X}(\pmb{q}^2)$ decomposed into logitudinal and transverse components of the channels $V^\dagger_\mu V_\mu $ and $A^\dagger_\mu A_\mu $.}
\label{fig:ContributionsXBar}
\end{figure}

In Fig. \ref{fig:ContributionsXBar} we show the contributions to the total $\bar{X}(\pmb{q}^2)$, which is defined from Eq. \eqref{equ:TotalDecayRate} as $\bar{X}(\pmb{q}^2) = \sum_{l=0}^{2} \bar{X}^{(l)}(\pmb{q}^2)$. We decompose $\bar{X}(\pmb{q}^2)$ into contributions of the vector and axial-vector channels, denoted by $VV$ and $AA$, respectively, in order to expose the $V-A$ nature of the charged currents. Contributions of the type $AV$ and $VA$ vanish in the case of massless leptons assumed throughout the analysis. We have further sub-divided the contributions into logitudinal and transversal components. This decomposition allows us to perform an important cross-check in our analysis, i.e. compare the results to the ones obtained under the ground-state limit. A detailed discussion can be found in \cite{Barone:2023tbl}, where we showed that the results show great agreement in the ground-state limit.

\section{Systematic errors in the inclusive analysis}
\label{sec:SystematicErrors}

We focus on two major sources of systematic errors in the inclusive decays, namely the error due to the polynomial approximation and the error due to finite volume effects. The former is a combination of two effects: first, the smoothing of the kernel function requires the $\sigma \rightarrow 0$ limit and secondly, the truncation of the Chebyshev approximation at polynomial order $N$ requires the $N\to\infty$ limit. We refer to \cite{Kellermann:2022mms}, where we present a method based on properties of the Chebyshev polynomials in order to obtain an estimate for the corrections expected from the two required limits.

\begin{figure}[hbt!]
  \centering
  \begin{subfigure}{0.49\textwidth}
    \centering
    \includegraphics[width=0.8\textwidth]{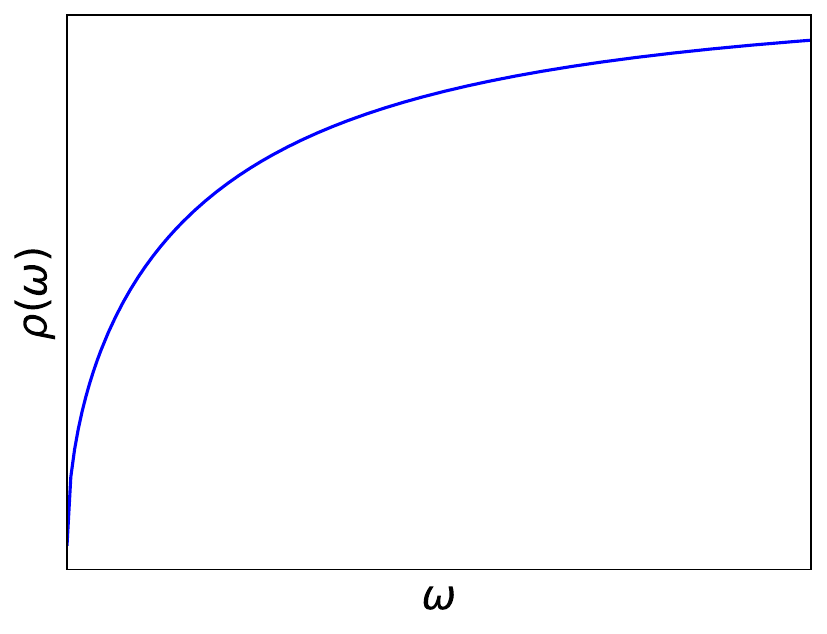}
  \end{subfigure}
  \begin{subfigure}{0.49\textwidth}
    \centering
    \includegraphics[width=0.8\textwidth]{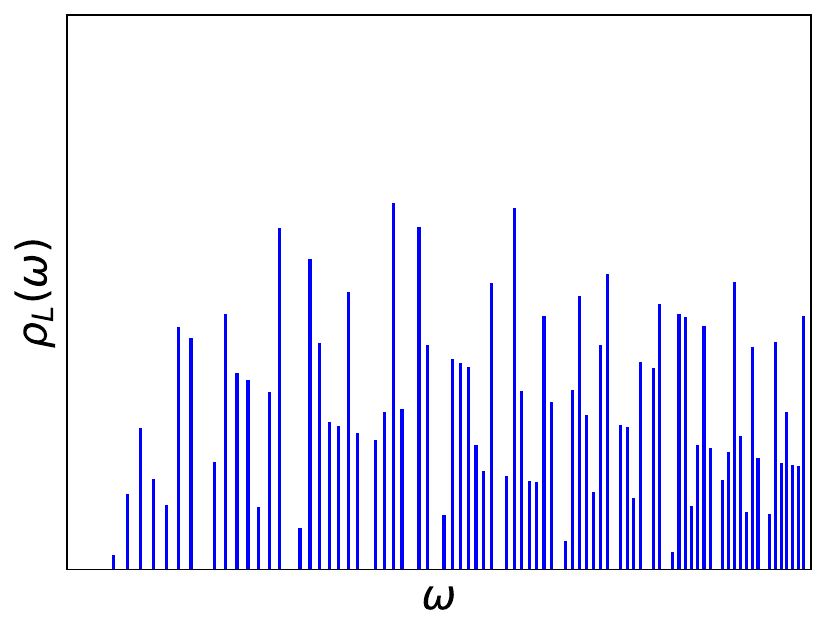}
  \end{subfigure}
  \caption{Sketch of the infinite-volume spectral density $\rho(\omega)$ (left) and the finite-volume $\rho_V(\omega)$ for a specific volume $V$ (right). The height of $\rho_V(\omega)$ represents the multiplicity of the states with the same energy $\omega$.}
  \label{fig:FiniteVolSPectralFunction}
\end{figure}

The error due to finite volume effects is a well known challenge for any calculations performed on the lattice and concerns the reconstruction of the spectral density from correlators $C(t)$ with a finite set of discrete time slices. This is commonly referred to as the ill-posed inverse problem. Even if the inverse problem could be solved for a correlator in a finite volume, $C_{V}(t)$, where $V=L^3$ denotes the volume of the lattice, and hence the spectral density $\rho_V(\omega)$ is reconstructed, there is still a qualitative difference from its infinite volume counterpart $\rho(\omega)$. The spectral density in the infinite volume is a smooth function, while $\rho_V(\omega)$ is given by a sum of $\delta$-peaks representing allowed states in a finite volume. In Fig. \ref{fig:FiniteVolSPectralFunction} we sketch the situation for two-body states in a finite volume, which are knwon to receive corrections of the order $\mathcal{O}(1/L^3)$.
This problem is avoided by the introduction of the smearing in the kernel function $K(\omega)$ as shown in Eq. \eqref{equ:KernelFunction}. The inverse problem is made arbitrarily mild by increasing the smearing width $\sigma$, and the smeared spectral density $\rho_{\sigma, V}$ then smoothly approaches its infinite volume counterpart. To recover the inclusive decay rate, we therefore need to take the limit $V\to\infty$ before taking the limit of vanishing smearing width. Conventionally, the infinite volume is estimated by extrapolating the results from the same calculation for different choices of finite volumes. We refer to \cite{Kellermann:2023yec} for a discussion on a modeling strategy based on the assumption of multi-body states in order to estimate finite-volume corrections for cases where the data for the conventional approach is not available.

\section{Summary and Future Prospects}
\label{sec:Future}

In this work, we have extended the idea presented in \cite{Gambino:2020crt} and applied it to the studies of the inclusive semileptonic decay of the charmed $D_s$ mesons from lattice correlators using a full and flexible lattice setup. For analyzing the inclusive process we employ the Chebyshev polynomials with a trade off between the data and the Chebyshev matrix elements to account for the bounds on the Chebyshev polynomials. In \cite{Barone:2023tbl}, we were able to confirm that the Chebyshev approximation enables an accurate and efficient way to determine the inclusive decay rate. A first comparison with experimental data taken from the BESIII collaboration indicates that our predictions are in the right ballpark, giving us confidence that this method will provide fruitful input in the future.

We also discussed methods towards estimating the systematic errors, although these have to be worked out in more detail in the future. As we concluded in \cite{Kellermann:2022mms}, while using the properties of the Chebyshev approximation is mathematically sound, this approach yields a very conservative estimate on the systematic error associated with the approximation. Similarily, in \cite{Kellermann:2023yec} we arrive at the conclusion that a deeper understanding on the dependence of the finite-volume effects on the choice of the model is imperative and that supplementing the model with more data is desirable in order to verify the validity of the model. In addition to the errors discussed here future studies will also need to focus on additional sources of systematic errors such as discretization errors or the continuum limit. Plans to address these are currently in development and the data generation will begin in the near future.

Taking a look towards additional future developments, there is on-going work towards applying the Chebyshev approximation towards different observables in the inclusive decays, e.g. moments such as $q^2$ or lepton energy moments \cite{Gambino:2022dvu, Barone:2023iat}. The basic analysis strategy can be applied for these observables. This analysis will increase the pool of predictions that can be compared to different theoretical predicitons obtained in the continuum theory, such as the operator-product expansion as well as experimental results.
Furthermore, following the preliminary study of the inclusive decay of $B_{(s)}$ mesons performed in \cite{Barone:2023tbl}, we plan to verify whether our methods towards estimating the systematic errors discussed in this work can straightforwardly be applied in the case of the bottom sector. The final goal is then to combine our findinds with experimental data in order to provide a new and independent determination of the CKM matrix elements for both the charm and bottom sector.

\section*{Acknowledgments}

The numerical calculations of the JLQCD collaboration were performed on SX-Aurora TSUBASA at the High Energy Accelerator Research Organization (KEK) under its Particle, Nuclear and Astrophysics Simulation Program, as well as on Fugaku through the HPCI System Research Project (Project ID: hp220056).

The works of S.H. and T.K. are supported in part by JSPS KAKENHI Grant Numbers 22H00138 and 21H01085, respectively, and by the Post-K and Fugaku supercomputer project through the Joint Institute for Computational Fundamental Science (JICFuS).



\begin{thebibliography}{99}

\bibitem{Workman:2022ynf}
R.~L.~Workman \textit{et al.} [Particle Data Group],
PTEP \textbf{2022} (2022), 083C01
doi:10.1093/ptep/ptac097

\bibitem{Gambino:2020crt}
P.~Gambino and S.~Hashimoto,
Phys. Rev. Lett. \textbf{125} (2020) no.3, 032001
doi:10.1103/PhysRevLett.125.032001
[arXiv:2005.13730 [hep-lat]].

\bibitem{Gambino:2022dvu}
P.~Gambino, S.~Hashimoto, S.~M\"achler, M.~Panero, F.~Sanfilippo, S.~Simula, A.~Smecca and N.~Tantalo,
JHEP \textbf{07} (2022), 083
doi:10.1007/JHEP07(2022)083
[arXiv:2203.11762 [hep-lat]].

\bibitem{Hansen:2017mnd}
M.~T.~Hansen, H.~B.~Meyer and D.~Robaina,
Phys. Rev. D \textbf{96} (2017) no.9, 094513
doi:10.1103/PhysRevD.96.094513
[arXiv:1704.08993 [hep-lat]].

\bibitem{Hansen:2019idp}
M.~Hansen, A.~Lupo and N.~Tantalo,
Phys. Rev. D \textbf{99} (2019) no.9, 094508
doi:10.1103/PhysRevD.99.094508
[arXiv:1903.06476 [hep-lat]].

\bibitem{Bulava:2021fre}
J.~Bulava, M.~T.~Hansen, M.~W.~Hansen, A.~Patella and N.~Tantalo,
JHEP \textbf{07} (2022), 034
doi:10.1007/JHEP07(2022)034
[arXiv:2111.12774 [hep-lat]].

\bibitem{Kellermann:2022mms}
R.~Kellermann, A.~Barone, S.~Hashimoto, A.~J\"uttner and T.~Kaneko,
PoS \textbf{LATTICE2022}, 414 (2023)
doi:10.22323/1.430.0414
[arXiv:2211.16830 [hep-lat]].

\bibitem{Barone:2023tbl}
A.~Barone, S.~Hashimoto, A.~J\"uttner, T.~Kaneko and R.~Kellermann,
JHEP \textbf{07}, 145 (2023)
doi:10.1007/JHEP07(2023)145
[arXiv:2305.14092 [hep-lat]].

\bibitem{Kellermann:2023yec}
R.~Kellermann, A.~Barone, S.~Hashimoto, A.~J\"uttner and T.~Kaneko,
[arXiv:2312.16442 [hep-lat]].

\bibitem{Barone:2023iat}
A.~Barone, S.~Hashimoto, A.~J\"uttner, T.~Kaneko and R.~Kellermann,
[arXiv:2312.17401 [hep-lat]].

\end{thebibliography}
\end{document}